\documentclass[aps, prb, 10pt, preprint, showpacs, preprintnumbers,
superscriptaddress, amsmath, amssymb]{revtex4-1}
\usepackage{graphicx}
\usepackage{epstopdf}
\usepackage{color}
\usepackage{xfrac}
\usepackage[utf8]{inputenc}

\begin{document}
\title{Electric field sensing with a scanning fiber-coupled quantum dot}
\author{D. Cadeddu} \affiliation{Department of Physics, University of Basel, 4056 Basel, Switzerland}
\author{M. Munsch} \affiliation{Department of Physics, University of Basel, 4056 Basel, Switzerland}
\author{N. Rossi} \affiliation{Department of Physics, University of Basel, 4056 Basel, Switzerland}
\author{J. Claudon} \affiliation{INAC-PHELIQS, ``Nanophysique et semiconducteurs'' group, CEA, Univ.\ Grenoble Alpes, 38054 Grenoble, France}
\author{J.-M. G\'erard} \affiliation{INAC-PHELIQS, ``Nanophysique et semiconducteurs'' group, CEA, Univ.\ Grenoble Alpes, 38054 Grenoble, France}
\author{R. J. Warburton} \affiliation{Department of Physics, University of Basel, 4056 Basel, Switzerland}
\author{M. Poggio} \affiliation{Department of Physics, University of Basel, 4056 Basel, Switzerland} \email{martino.poggio@unibas.ch} \homepage{http://poggiolab.unibas.ch/}

\date{\today}

\begin{abstract}
We demonstrate the application of a fiber-coupled quantum-dot-in-a-tip as a probe for scanning electric field microscopy. We map the out-of-plane component of the electric field induced by a pair of electrodes by measurement of the quantum-confined Stark effect induced on a quantum dot spectral line. Our results are in agreement with finite element simulations of the experiment. Furthermore, we present results from analytic calculations and simulations which are relevant to any electric field sensor embedded in a dielectric tip. In particular, we highlight the impact of the tip geometry on both the resolution and sensitivity.
\end{abstract}
\maketitle

Motivated by a desire to measure the electronic properties of surfaces and nano-objects, many nanoscale electric field sensors have been developed over the years. Electrostatic force microscopy (EFM)\cite{martin_highresolution_1988}$^,$\cite{gross_measuring_2009}, scanning Kelvin probe force microscopy (KPFM)\cite{henning_twodimensional_1995}, and scanning single electron transistors (SETs)\cite{yoo_scanning_1997}$^,$\cite{martin_observation_2008} have already established themselves as sensitive electric field detectors reaching sub-elementary charge sensitivity and sub-nanometer spatial resolution. More recently, single defect centers in diamond were also used to detect AC electric fields at room temperature\cite{dolde_electric-field_2011} and a scanning tunneling microscope (STM) functionalized with a single molecule was used to image the dipole field of an ad-atom on a surface\cite{wagner_scanning_2015}. In transport experiments, gate-defined quantum dots (QDs) have been employed as single charge detectors \cite{barthel_fast_2010}$^,$\cite{arnold_cavity_2014} and self-assembled QDs have been employed as all-optical electrometers, demonstrating a sensitivity of 5 $\text{V}/(\text{m} \cdot \sqrt{\text{Hz}})$ \cite{vamivakas_nanoscale_2011}. These kind of QDs were also used to determine the position of single defect charges within 100 nm of a QD with a precision of 5 nm \cite{houel_probing_2012}. Electric fields in QDs produce large Stark shifts, which, due to a built-in electric dipole, are nearly linear around zero field. Nevertheless, a scanning electric field sensor based on an optically active semiconductor QD has not yet been realized. Such a scanning probe has the potential for a very large bandwidth, which unlike electronic and mechanically addressable sensors, is limited only by the spontaneous emission rate of the QD and could therefore approach the GHz range.

Here we show the proof-of-principle application of an optical fiber-coupled semiconductor QD as a scanning electric field sensor. By tracking the induced energy shift on the peak of a single transition of a QD, we are able to map the vertical component of an external applied field. Our device is composed of a self-assembled InAs QD located in the tip of a fiber-coupled GaAs photonic wire. The QD emits preferentially into the waveguide mode, which expands adiabatically through a tapering of the photonic wire and ensures  good out-coupling\cite{munsch_dielectric_2013}. The design of the photonic wire ensures both efficient guiding of the QD fluorescence as well as proximity of the QD to the sample, which is a necessary condition for scanning probe microscopy.  With the help of numerical simulations, we also show the unavoidable perturbation of the external field due to the dielectric nature of the probe, pointing toward geometric improvements to reduce this effect and increase the sensitivity of the device.

\begin{figure}[t]
\centering
\includegraphics[width=1\textwidth]{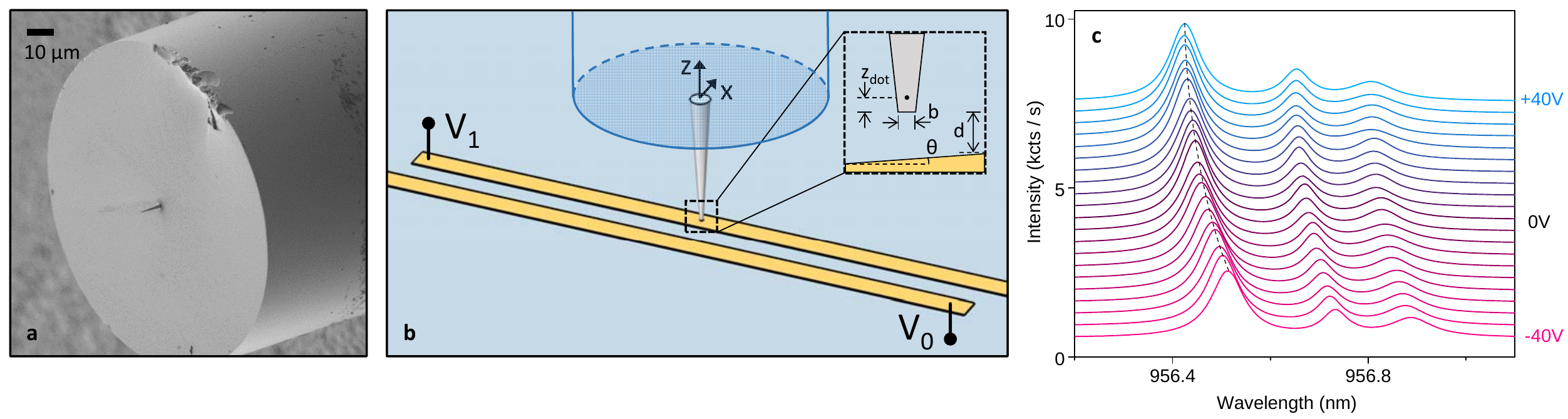}
\caption{Fiber-coupled QD electrometer - (a) Scanning electron micrograph (SEM) of the fiber-coupled photonic wire. (b) Schematic of the measurement setup. Inset: cross section of the $xz$-plane. Tip-sample distance $d$, tip bottom diameter $b$, QD distance from the base of the tip $z_{dot}$, and sample tilt angle $\theta$. (c) Lorentzian fits of the investigated part of the PL spectrum showing excitonic lines associated with one QD for different values of $V_1$. The graphs are offset for clarity.}
\label{fig:figure1}
\end{figure}

The photonic trumpet used here is obtained through an etching process carried out on a GaAs wafer containing a layer of self-assembled InAs QDs grown by molecular beam epitaxy \cite{munsch_dielectric_2013}. The layer of QDs is located at $z_{dot}=110$~nm above the bottom facet. The Gaussian intensity profile obtained at the top facet of the 11-$\mu$m-tall wire structure enhances the coupling of the QD emission directly into the single-mode fiber~\cite{stepanov_highly_2015}. We select a wire with a top diameter of 1.64~$\mu$m, and a bottom diameter $b=350$~nm. The wire is then cleaved from the substrate under an optical microscope using a micro-manipulator. Using the same apparatus, it is then transferred and glued to the core of a single-mode optical fiber. Optimal coupling is obtained by centering the wide top facet of the photonic wire onto the core of the fiber\cite{cadeddu_fiber-coupled_2016}, as shown in Fig.~\ref{fig:figure1}a. The final device constitutes a reliable fiber-coupled source of single photons with a photon collection efficiency of $\sim 6\%$, which is an order of magnitude better than directly coupling the fiber to a QD in bulk GaAs. 


In order to investigate the performance of our probe as a sensor of electric field, we mount it in a low-temperature scanning probe microscope. The QDs at the end of the photonic wire are excited non-resonantly with a CW laser diode at 830~nm, which  excites carriers directly in the wetting layer and avoids heating of the GaAs wire~\cite{cadeddu_fiber-coupled_2016}. Photoluminescence (PL) from the QDs is guided into the fiber by the photonic wire and analyzed with a spectrometer equipped with a CCD camera. The fiber-coupled photonic tip is then positioned over a pair of parallel Au electrodes, across which we apply a voltage $V_1 - V_0$, with $V_0$ fixed to ground. The tip is positioned about 10 nm above the gate, held at potential $V_1$. The electrodes, shown schematically in Fig.~\ref{fig:figure1}~b), are deposited on a Si/SiO$_2$ substrate and are 80-nm-thick, 2-$\mu$m-wide, and 2-$\mu$m-apart from each other. The resulting electric field tilts the energy bands of the semiconductor and, due to the quantum-confined Stark effect~\cite{fry_inverted_2000}$^,$\cite{schulhauser_magneto_2002}, the transition energies of each QD are shifted to lower energy. Here we focus on a bright peak centered at 956.4 nm with a line-width of $100~\mu$eV, which we attribute to an excitonic transition in a single QD. By sweeping the applied voltage $V_1$ from $-40$  to $+40$ V we observe a shift in the energy of the emitted PL that is well-described by a quadratic function of the applied voltage -- and therefore of the applied electric field -- as shown in Fig.~\ref{fig:figure1}~c):
\begin{equation}
\centering
\xi=\xi_0 - p_\parallel E_\parallel - p_\perp E_\perp + \beta_\parallel E_\parallel^2 + \beta_\perp E_\perp^2,
\label{eq:Stark}
\end{equation}
\noindent where $\xi_0$ is the unperturbed energy; $p_\parallel$ ($p_\perp$) is the static electric dipole of the QD exciton parallel (perpendicular) to the wire axis; and $\beta_\parallel$ ($\beta_\perp$) is the polarizability  of the QD exciton parallel (perpendicular) to the wire axis. The electric field at the QD location $\vec{r}$ is the sum of two components $\vec{E}(\vec{r}) + \delta \vec{E}(\vec{r})$, where the first term is due to the field applied across the electrodes and the second term is due to charges trapped in the vicinity of the QD. $\vec{E}(\vec{r}) = \vec{\alpha}(\vec{r}) (V_1 - V_0)$, where $\vec{\alpha}(\vec{r})$ is a position dependent proportionality constant, which describes the spatial configuration of the the electric field produced by the split-gates and the dielectric wire. $\delta \vec{E}(\vec{r})$ describes a random electric field that changes upon the reorganization of charges near the QD.

\begin{figure}[t]
\centering
\includegraphics[width=0.8\textwidth]{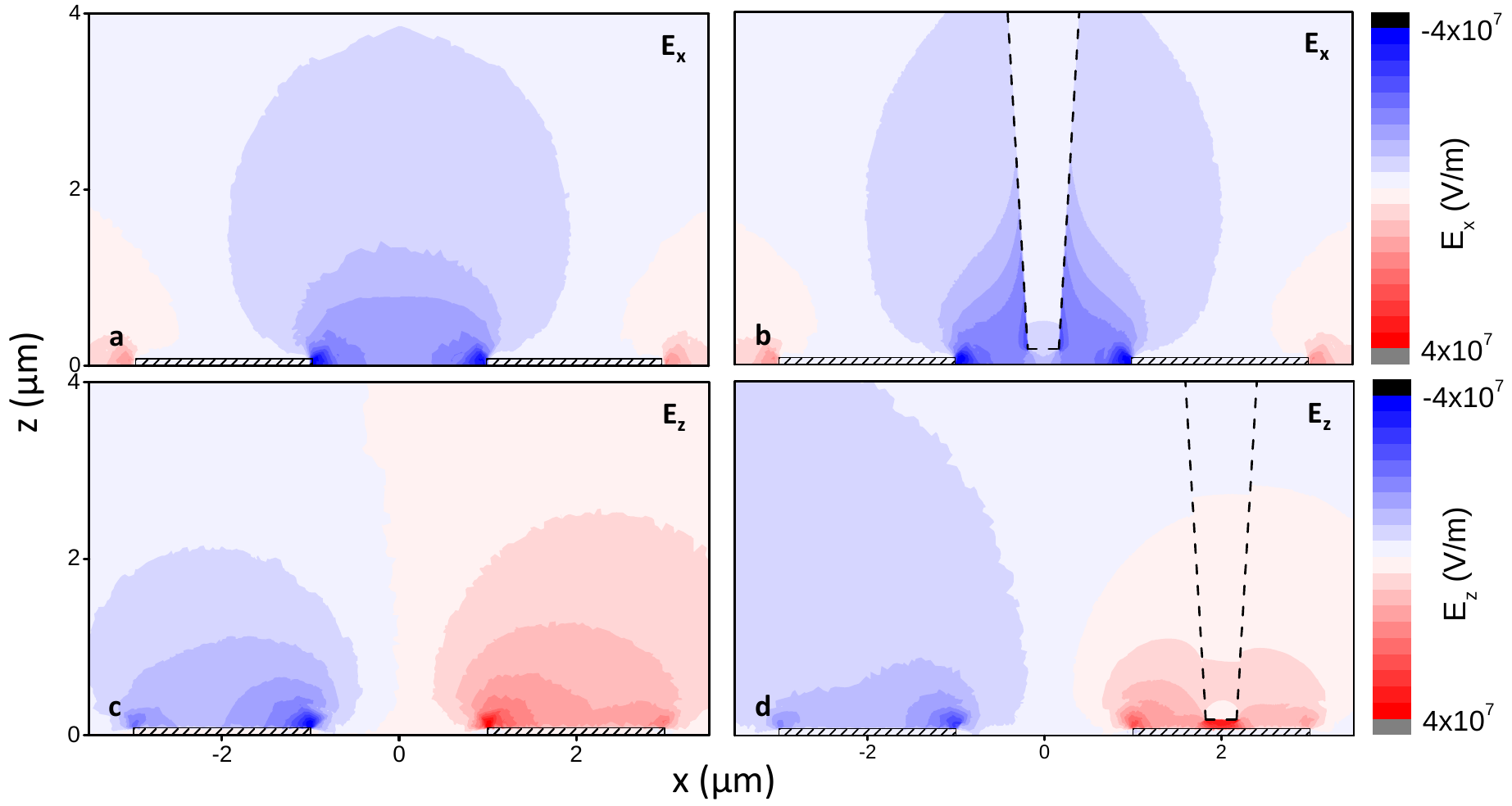}
\caption{Simulated maps of the $\vec{E}(\vec{r})$ - The components (a) $E_x$ and (c) $E_z$ generated by the pair of electrodes set to $V_1=40$ and $V_0=0$~V., (b) and (d) show the same electric field components  in the presence of the dielectric tip. Note the pronounced screening of $E_x$ by the photonic tip compared to that of $E_z$ at the position of the QD near the apex of the probe. }
\label{fig:figure2}
\end{figure}

It is important to note that the presence of the photonic wire, due to its dielectric nature, reduces the field at the QD position and significantly perturbs the external applied field. In the simplest approximation, a thin dielectric cylinder with a uniform and unidirectional polarization strongly suppresses electric fields applied perpendicular to its long axis, while leaving parallel fields largely unchanged. An analytic model of a conical section matching our photonic trumpet geometry, as well as a more realistic finite element calculation, show that the wire's narrow radial cross-section results in an efficient screening of $E_x$, while $E_z$ is less affected (see Supporting Information). At the QD position $z_{dot}=110$~nm above the bottom facet, this directional screening effectively projects the unperturbed electric field along $\hat{z}$. The magnitude of the effect depends on the diameter of the facets, the length of the wire, the position of the QD, and the direction of the field with respect to the long axis of the wire. This screening, as well as a distortion of the electric field outside the photonic tip, are shown in simulations of our experimental electrode geometry in Fig.~\ref{fig:figure2}. In our experiment, the field at the dot position reaches values of $60\%$ and $20\%$ of the applied external field in the $\hat{z}$ and $\hat{x}$ direction, respectively.

We map the spatial dependence of the electric field produced by the gates by scanning the photonic wire tip and sweeping the applied voltage at every position. We scan above one electrode, where the electric field points nearly exclusively along $\hat{z}$. Given the direction and the preferential penetration of the field in the $\hat{z}$ direction, we can assume $E_x$ to be negligible in this region. $E_y$ vanishes due to the symmetry of the electrode structure.  Given that $E_\parallel = E_z + \delta E_z$ and $E_\perp = \sqrt{(E_x + \delta E_x)^2+(E_y + \delta E_y)^2}$, (\ref{eq:Stark}) becomes:
\begin{equation}\label{eq:Stark2}
\begin{split}
\xi &= \Big [\xi_0 - p_\parallel \delta E_z - p_\perp \sqrt{\delta E_x^2 + \delta E_y^2} + \beta_\parallel \delta E_z^2 + \beta_\perp \left (\delta E_x^2 + \delta E_y^2 \right ) \Big ] \\
&  \qquad - \Big [p_\parallel - 2 \beta_\parallel \delta E_z \Big ] E_z + \beta_\parallel E_z^2 
\end{split}
\end{equation}
\noindent Note that both the constant and linear terms in $E_z$ depend on the random electric field due to charging effects. The quadratic term, on the other hand, depends only on the polarizability along the wire axis.

\begin{figure}[t]
\centering
\includegraphics[width=1\textwidth]{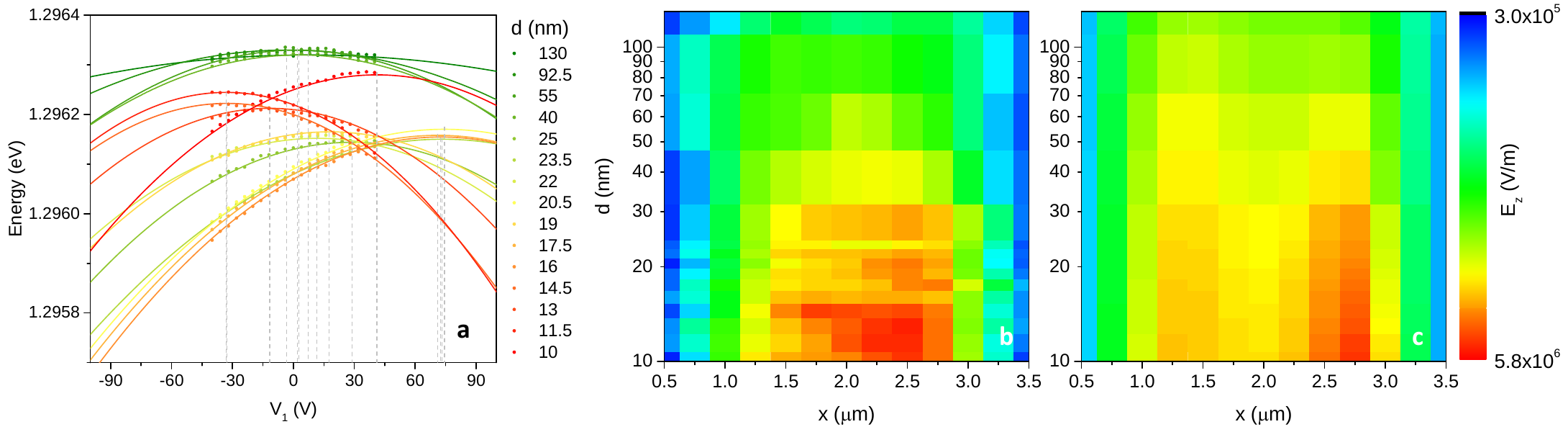}
\caption{Experimental results - (a) Energy of the leftmost peak in Fig.~\ref{fig:figure1}(d) versus the applied voltage $V_1$ for different values of tip-electrode distance $d$. Solid lines are extended parabolic fits for different values of $d$ and dashed gray lines indicate the vertices of each parabola. The tip is positioned at the center of the electrode ($x=2$ $\mu$m).
(b) Map of $E_z$ at $V_1-V_0=40$ V extracted from measurements of QD emission energy as a function of position in the $xz$-plane. (c) Simulated values of $E_z$ at the QD position $z_{dot}=110$~nm for  a photonic tip with $b=350$~nm and $V_1-V_0=40$ V. The tilt angle $\theta$ is set to $5$ degrees to match qualitatively the experimental data.}
\label{fig:figure3}
\end{figure}

By collecting PL spectra, we measure the dependence of $\xi$ on both voltage and position in the $xz$-plane. The measured QD exciton energies show a parabolic dependence on $V_1$ with an offset $a_0$, a linear coefficient $a_1$, and a curvature $a_2$, each depending on position in the $xz$-plane, as seen in Fig.~\ref{fig:figure3}~a). As expected from the dependence of the constant and linear terms in (\ref{eq:Stark2}) on components of $\delta \vec{E}$, $a_0$ and $a_1$ appear random and are observed to be hysteretic in both voltage and position. They are likely determined by the charging and discharging of defects within the photonic wire, which generate an extra electric field in the vicinity of the QD. On the other hand, $a_2$ remains constant as a function of voltage and reproducible as a function of position, following what is expected from (\ref{eq:Stark2}): $a_2(x,z) = \beta_\parallel \alpha_z (x,z)^2$, where $\beta_\parallel$ is a constant and $\alpha_z(x,z)$ is set by the configuration of the electrodes.

In order to make a detailed comparison to the experiment, we make a finite element simulation of $\alpha_z(x,z)$ at the position of the QD as the photonic wire is scanned above the electrode.  A corresponding experimental map of this term can be extracted from the fits to the measured data, since $\alpha_z(x,z) = \sqrt{a_2(x,z)/\beta_\parallel}$.  In Figs.~\ref{fig:figure3}~b) and c), we plot the measured and simulated $E_z(x,z)$, respectively, corresponding to an applied voltage of $V_1-V_0=40$ V. In order to match the spatial dependence of our measurements to the simulations, we introduce a tilt angle $\theta = 5~^\circ$, as shown in the inset of Figure~\ref{fig:figure1}b. Such a misalignment between both scanning stages is experimentally reasonable and, in practice, difficult to avoid. A polarizability $\beta_\parallel=-0.012 \pm 0.005~\mu$eV/(kV/cm)$^2$ brings the measured and simulated values of $E_z(x,z)$ into numerical agreement, as shown in Fig.~\ref{fig:figure3}~b) and c). This measured polarizability is an order of magnitude smaller than what is typically observed in literature for single exciton transitions in similar QDs\cite{warburton_giant_2002}$^,$\cite{prechtel_electrically_2015}$^,$\cite{finley_quantum-confined_2004}.
Nevertheless, the similarity of the measured and simulated maps of $E_z(x,z)$, demonstrates the successful implementation of our QD probe to spatially map the magnitude of a DC electric field along one direction. Note that despite the fact that we require fits to full voltage sweeps at each position to extract the electric field configuration, applying a small AC voltage should simplify and speed up the measurement. The AC response at $f$ and $2 f$, where $f$ is the frequency of the applied field, would allow separation of the linear contribution, arising from the electric dipole of the QD, from the quadratic contribution, arising from the polarizability, respectively~\cite{Rossi_vectorial_2017}. 

\begin{figure}[t]
\centering
\includegraphics[width=0.9\textwidth]{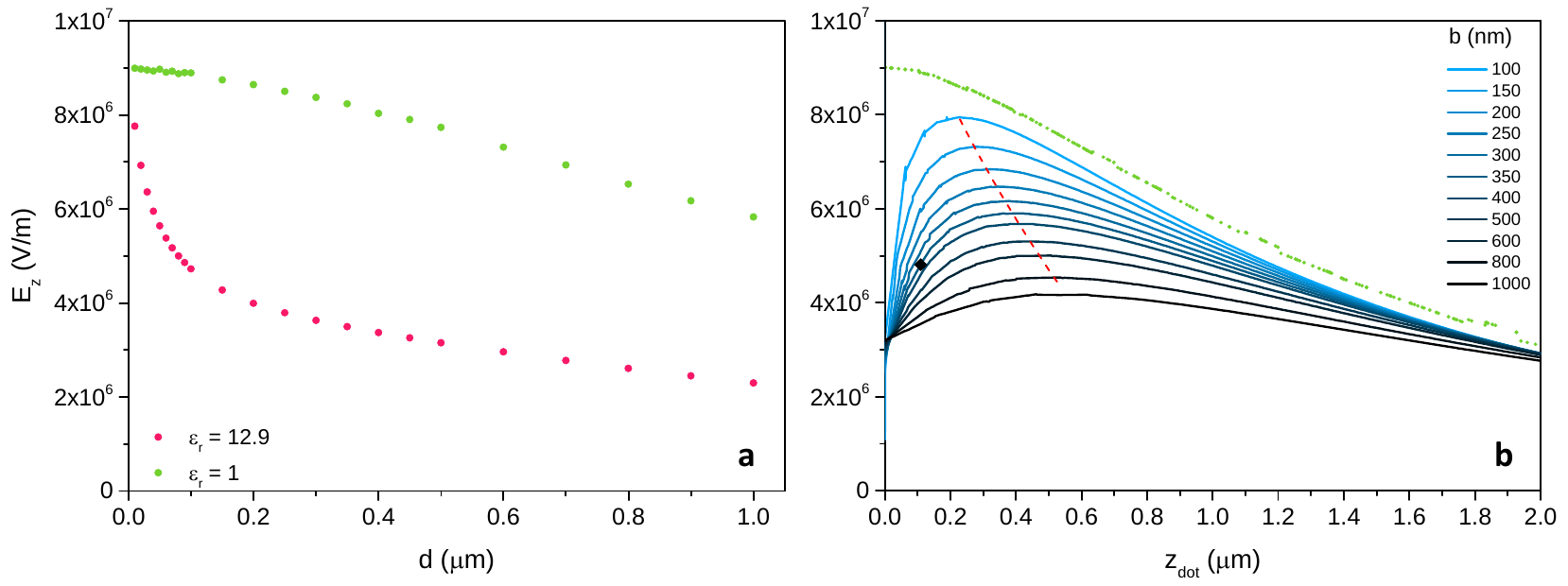}
\caption{Influence of the tip geometry on sensing capabilities - (a) Simulated $E_z$ at $z_{dot}=110~nm$ is plotted in pink as a function of tip-electrode distance $d$ in a GaAs tip with $b=350~$nm. The corresponding  condition without screening is plotted in green. (b) $E_z$ values along a $z$ line-cut with the photonic tip at $d=0.1~\mu m$ for different values of the bottom diameter $b$. We define $z_{opt}$ as the position where $E_z$ is maximum inside the tip. The values of $z_{opt}$ are fitted by the red dashed line for different values of $b$. The black diamond indicates the QD position in our current device, while the green dots are values of $E_z$ in the absence of the trumpet, as a reference.}
\label{fig:figure4}
\end{figure}

One important observation that can be made from Fig.~\ref{fig:figure3}~b) and c) is the rapid decrease in measured electric field as a function of tip-electrode distance $d$ compared to that expected in vacuum. As shown in Fig.~\ref{fig:figure4}~a), this effect is a direct consequence of the polarization charge induced on the bottom facet of the photonic wire, which screens the out-of-plane electric field impinging on the QD. Note that this effect must be considered for any electric field sensor based on a dielectric scanning probe. In order to reduce the screening effect for $E_z$, which decreases the sensitivity of the sensor and distorts the observed field with respect to the unperturbed case, this surface charge density must be minimized. If we exclude replacing the GaAs tip material with one having a dielectric constant closer to the one of the vacuum, the reduction of this screening charge can be achieved by reducing the diameter $b$ of the bottom facet. This sharpening of the scanning probe, would also help to increase the screening of $E_x$, making the QD Stark shift an even closer measure of ${E_z}^2$. In practice, however, $b$ is constrained to values higher than $190$ nm in order to maintain the guiding of the QD PL up the photonic wire and into the optical fiber above $90\%$ \cite{friedler_solid-state_2009}. The choice of $b$ in turn sets a natural minimum tip-sample distance $d$ of the same order, below which both the spatial resolution will not improve and the tip will strongly perturb the local electric field. Once $b$ and $d$ are fixed, finite element calculations show that there exists an optimal position for the QD above the bottom facet $z_{opt}$, as shown by Fig.~\ref{fig:figure4}~b). This position minimizes the screening effect and hosts the largest measurable field from the electrodes within the tip. The black diamond in Fig.~\ref{fig:figure4}~b) indicates the position of the QD within the device used in our experiments. For $d = 100$~nm, optimization of $b$ and of the position of the QD could result in the ability to measure much weaker fields parallel to the wire axis with almost no tip-induced perturbation of this component. Note that in the case of optical readout, $z_{dot}$ is bound to be at multiple of $\lambda/2n$ from the bottom facet, where $\lambda$ is the wavelength of the emitted light and $n$ the index of refraction of the wire, in order to ensure maximum reflection in the upward direction. In general, there is a natural trade off between a scanning probe tip which minimally perturbs the electric field and one that optimally guides the sensor emission.

In conclusion, we demonstrate the application of a fiber-coupled semiconductor QD as a scanning probe for electric field sensing. The probe QD resides at the apex of a sharp GaAs scanning probe, which allows both for an efficient optical collection and a preferential penetration of on-axis electric fields. Despite position-dependent hysteresis and randomness exhibited by the QD's PL energy and electric dipole, its constant polarizability allows for the extraction of the electric field magnitude at the position of the probe QD. By scanning this sensing element, we map the spatial dependence of the out-of-plane electric field produced by a pair of micro-fabricated electrodes at 4~K. By optimizing the geometry of the photonic wire tip and the position of the QD, higher optical collection efficiencies could be combined with better electric field penetration. Such improvements should lead to higher sensitivity and reduced distortion of the unperturbed fields. Furthermore, measuring by resonance fluorescence should reduce the line-width of the QD and by consequence enhance the precision of the measurements. Implementing such improvements could lead to an all-optical scanning electrometer with GHz bandwidth and sensitivities of 20 $\text{V}/(\text{m} \cdot \sqrt{\text{Hz}})$~\cite{cadeddu_fiber-coupled_2016}. Such sensitivity, combined with a spatial resolution roughly set by the bottom diameter of the tip would allow for the mapping of single charges on surfaces and may provide an alternative to scanning single electron transistors.

\begin{acknowledgments}
We thank Sascha Martin for technical support. This work is supported by an ERC Grant (NWscan, Grant No. 334767), the Swiss Nanoscience Institute (Project P1207), and the National Centre of Competence in Research, Quantum Science and Technology. Sample fabrication was carried out in the Upstream Nanofabrication Facility (PTA) and CEA LETI MINATEC/DOPT clean rooms.
\end{acknowledgments}

%

\end{document}